\documentclass[11pt]{article}         
\usepackage {amsfonts,citesort} 

\begin{document}
\title{Null Cones and Einstein's Equations in Minkowski Spacetime}

\author{J. Brian Pitts\footnote{The Ilya Prigogine Center for Studies in
     Statistical Mechanics and Complex Systems, 
Department of Physics,
     The University of Texas at Austin,
	Austin, Texas 78712; 
	Mathematics Department, 
	St. Edward's University,
	 Austin, Texas 78704; 
current address:  Program in 	History and Philosophy of Science,
346 O'Shaughnessy, 
University of Notre Dame, 
Notre Dame, Indiana 46556; 
email jpitts@nd.edu    \vspace{2 mm}  }
  and W. C. Schieve\footnote{The Ilya Prigogine Center for Studies in
     Statistical Mechanics and Complex Systems, 
     The University of Texas at Austin,
	Austin, Texas 78712} }

\vspace{.5in} 

\maketitle


\begin{abstract}

	If Einstein's equations are to describe a field theory of gravity in Minkowski spacetime, then causality requires that the effective curved metric must respect the flat background metric's null cone.  The kinematical problem is solved using a generalized eigenvector formalism based on the Segr\'{e} classification of symmetric rank 2 tensors with respect to a Lorentzian metric.  Securing the correct relationship between the two null cones dynamically plausibly is achieved using the naive gauge freedom.  New variables tied to the generalized eigenvector formalism reduce the configuration space to the causality-respecting part.  In this smaller space, gauge transformations do not form a group, but only a groupoid.  The flat metric removes the difficulty of defining equal-time commutation relations in quantum gravity and guarantees global hyperbolicity.
\end{abstract}


key words:  null cones, bimetric, spin 2, field formulation, causality



\section{Introduction}                      

     A number of authors have discussed the utility of a flat
background metric $\eta_{\mu\nu}$ in general relativity.  Remarkably, a background metric  also enables one to \emph{derive} general relativity from plausible postulates in special relativity  \cite{GuptaReview,Kraichnan,Thirring,NSSexlField,Huggins,Feynman,Weinberg64c,Weinberg64d,WyssGC,OP,vanNieuwenhuizen,Deser,DeserQG,Fronsdal,Cavalleri,DaviesFang,Meszaros,Grishchuk,Grishchuk90,LogunovFund,LogunovBook,SliBimGRG,BoulangerEsole}. It is worth recalling a conclusion of E. R. Huggins \cite{Huggins}, who was a student of Feynman.  Huggins found that the
 requirement that energy be a spin-two field coupled to the stress-energy tensor does not lead to a unique theory.  Rather, ``an additional restriction is necessary.  For Feynman
this restriction was that the equations of motion 
be obtained from an action principle; Einstein required that the gravitational field have a geometrical interpretation.  Feynman showed these two restrictions to be equivalent.'' \cite{Huggins} (p. 3)  
As W. Thirring observed, it is not clear \emph{a priori} why
(pseudo-) Riemannian geometry is to be preferred over all the other sorts of geometry that exist, so a derivation of effective (pseudo-) Riemannian geometry is attractive \cite{Thirring}.  This remark is perhaps even more significant today, when  metric-affine geometries deprive metric geometry of much claim to generality or necessity.

	Such derivations of general relativity from Minkowski spacetime as exist to date, however, are only \emph{formally} special relativistic, because the null cone for the curved metric $g_{\mu\nu}$ might not respect the null cone for the flat metric $\eta_{\mu\nu}$.  The desired relationship is to have  all $\eta$-null vectors be $g$-null or $g$-spacelike, and all $\eta$-spacelike vectors be $g$-spacelike. Failing that, gravity would be an acausal theory in the sense of a field theory in Minkowksi spacetime.  Elsewhere we survey in some detail the history of the
treatment of this fundamental question \cite{NullCones,SliBimGRG}, and find that it remains open.
	
 	Here we undertake to solve the problem.  The kinematic issue of the relationship between the two null cones is handled using the work of G. S. Hall and collaborators on the Segr\'{e} classification of symmetric rank 2 tensors with respect
 to a Lorentzian metric.  For our purposes, we classify the curved metric with respect to the flat one, and find necessary and sufficient conditions for a suitable relationship. Requiring that flat spacetime causality not be violated, and not be arbitrarily close to being violated, a condition that we
 call ``stable $\eta$-causality,'' implies that all suitable curved metrics have a complete set of generalized eigenvectors with respect to the flat metric, and that the causality conditions take the form of strict rather than loose inequalities. Given strict inequalities, one is in a position to solve such conditions by a change of variables.  This technique is analogous to one used to satisfy
 the positivity conditions of canonical gravity, which have been discussed by J. Klauder, F. Klotz, and J. Goldberg.  
In these new variables, stable $\eta$-causality holds \emph{identically}, because the configuration space has been reduced (though the dimension is unchanged), largely by reducing the lapse until the proper null cone relation holds.  This reduction implies the need for reconsidering the gauge freedom of the theory.  It turns out that (suitably redefined) gauge transformations no longer form a group, because multiplication is not defined between some elements.  But they do form a groupoid, which seems quite satisfactory.  Given the plausibly successful outcome of the effort to make the proper null cone relationship hold, the above-mentioned gap in derivations of general relativity as an ostensibly special relativistic theory is substantially closed.  The naive gauge freedom turns out to include some unphysical states, but that is not a serious problem.  The goal of interpreting Einstein's equations in Minkowski spacetime is hardly new, but it seems never to have been successfully realized.  Some details of our approach are sufficiently novel that a new name is helpful to distinguish it from more formal and instrumental uses of a flat background metric tensor.  Let us call the work the ``special relativistic approach to Einstein's equations'' (SRA).  

	A natural use of the background causal structure is in defining equal time commutation relations in traditional approaches to quantum gravity.  Without a background causal structure of some sort, one cannot answer the kinematical question of which events occur at equal times until the dynamical problem has been solved.  Furthermore, because the curved metric is itself quantized, any resulting notion of causality will likely be fuzzy.  These problems are generated by the insistence on a fundamentally geometrical theory of gravity.  The background causal structure in the SRA naturally resolves these problems.

	Making the curved metric respect the flat null cone ensures that the resulting spacetime is globally hyperbolic.  This fact is quite consistent with the existence of a region of no escape, which can arise due to the narrowing and inward tilting of the curved null cones \cite{PetrovHarmonic}.  The mandatory global hyperbolicity might be useful in avoiding the Hawking black hole information loss paradox \cite{EarmanInfo}.  The background metric also renders the initial singularity of Robertson-Walker models innocuous, at least in some cases \cite{IARD2002}.   

\section{Describing and Enforcing the Proper Null Cone Relationship}

\subsection{Consistent Null Cones by Suitable Gauge Restrictions?}

     In pondering the relation between the local relation between the two null cones in a field formulation of general relativity, Penrose \cite{Penrose} and Grishchuk \cite{Grishchuk} have noted that it is gauge-dependent.  (Below we will modify the notion of gauge transformation to make the null cone relation gauge-invariant.)  This situation sounds problematic, but in fact it is not.  It is surely better than having the wrong null cone relationship in a physically invariant way, for example, as might happen in a bimetric theory with no gauge freedom. In fact, the gauge variance of the null cone relationship in a field approach to Einstein's equations is very helpful, because one is free to confine attention to those solutions with the proper null cone relationship, and then declare the remainder to be illegal.

	Looking at linearized general relativity helps to show both the prospects for getting the right null cone relationship, and an obstacle along the path.  Working in the Lorentz-Hilbert gauge, and excluding gravitational radiation, M. Visser, B. Bassett, and S. Liberati find that the curved metric's null cone opens wider than the flat background's only if the null energy condition (NEC) is violated 
\cite{SuperCens,PertSuperCens} (see also \cite{PenroseWoolgar,GaoWald,PalmerMarolf} for related work).  The NEC being rather commonly satisfied, and the Lorentz-Hilbert gauge being perhaps the most charming gauge condition, this result is quite encouraging for efforts to respect the flat metric's null cone in full nonlinear general relativity.  However, as they note, the NEC does not hold universally.  They also note that this result does not obviously or easily generalize to strong field situations.  Finally, we note that their exclusion of gravitational radiation is a severe limitation, for, as the study of gravitational waves in linearized general relativity shows, gravitational radiation in this gauge \emph{does} result in widening of the null cone relative to that of the flat background \cite{NullCones,Ohanian}.  The Lorentz-Hilbert gauge being the linearization of the  DeDonder gauge that holds when harmonic coordinates are used, its failure for plane wave solutions suggest that gauge fixing cannot be performed without due attention to the null cones.  This is our primary objection to the treatment of the two null cones in the Relativistic Theory of Gravity of A. A. Logunov \emph{et al.} \cite{LogunovFund,LogunovBasic,LogunovBook}.  Imposing a gauge condition up front and then discarding solutions with the wrong null cone relationship risks omitting physically necessary solutions, such as plane waves.  (We also reject Logunov's criticisms of general relativity as lacking conservation laws and failing to make unique physical predictions.)  Thus we concluded that causality is violated by physically relevant gravitational wave solutions in the tensorial DeDonder gauge \cite{NullCones,SliBimGRG}.

	In response to this argument, Yu. M. Chugreev has recently pointed out that in cases of demonstrably physically realistic gravitational radiation, one should not neglect the static field of the source, and that the static source field might prevent the gravitational wave from violating causality \cite{ChugreevCause} (see also (\cite{LoskutovRad96})).  For the finite-range massive graviton theory that Chugreev considers, if one considers an isolated source (not an unbounded cosmological matter distribution), one expects the radiation potential to fall off as $\frac{1}{r},$ while the static field suffers exponential decay from the Yukawa form.  Thus, at distances on the order of several Compton wavelengths, the source's static field will become negligible.  Then the linear gravitational wave solution alone would be physically relevant, and it would violate causality.  On the other hand, generalizing Chugreev's argument to the case of massless gravitons, which we consider here, both the radiation and the static field from a localized source should decay with distance as $\frac{1}{r}.$  Thus the radiation should not ``outlast'' the field of the source.  The static field might therefore prevent the gravitational wave from ever violating causality.  (These arguments have been made using a linear approximation.)  Whether there is causality violation from realistic exact solutions (expressed in the tensorial DeDonder gauge) describing the emission of gravitational radiation from localized sources is therefore presently an open question-- not settled either negatively was previously concluded in (\cite{NullCones,SliBimGRG}), or positively as some authors seem to believe.  In view of the difficulty in finding exact solutions of Einstein's equations, the services of numerical relativists--especially those who run simulations using harmonic coordinates--in running strong-field gravitational simulations would be helpful, and might be required, to resolve this issue.  In any case, it has become clear that gravitational wave solutions pose a larger question for the consistency of the null cones than do solutions with trivial or more monotonic dynamics.  For the present we conjecture that some realistic strong-field solution containing gravitational radiation, expressed using the tensorial DeDonder gauge, violates causality with respect to the flat background.  If this conjecture is correct, then our technique of using the gauge freedom to secure the proper relationship between the two null cones, to be discussed below, will be mandatory.  If our conjecture fails, then imposing the tensorial DeDonder condition ``by hand,'' as in the massless version of the Relativistic Theory of Gravity, would be an option.  However, if one wished to retain the gauge freedom that the constrained nature of the Einstein field equations naturally implies, then our technique below would still be required.  

	If there to be any is hope for restricting the gauge freedom so as to ensure that the curved null cone stays consistent with the flat one, then there must be enough naive gauge freedom to transform any
 physically significant solution into a form that satisfies $\eta$-causality. The curved spatial metric contains the gravitational degrees of freedom and satisfies constraint equations, so it is not readily adjusted.  Let us regard it as fixed, and attempt to use the gauge freedom resident in the lapse and shift \cite{MTW} to achieve null cone consistency.  Adjusting the lapse narrows or widens the curved metric's light cone, while the shift vector determines its tilt from the vertical (future) direction.  
By analogy with conditions typically imposed in geometrical general
relativity to avoid causality difficulties \cite{Wald}, one would prefer 
that the curved light cone be \emph{strictly inside} the flat light cone, not tangent to it, because tangency indicates that the field is on the verge of $\eta$-causality violation. Above we named this requirement ``stable $\eta$-causality,'' by analogy to the usual condition of stable causality \cite{Wald}.  One might worry that this requirement would exclude all curved metrics conformally related to the flat one,
 and even the presumed ``vacuum'' $g_{\mu\nu}=\eta_{\mu\nu}$ itself.  Indeed it does, but 
if we take gauge invariance seriously, then there is no fundamental basis for preferring $g_{\mu\nu}=\eta_{\mu\nu}$ over having the curved metric agree with the flat metric up to a gauge transformation.  Let the desired relation between the null cones hold at some initial moment.  Also let the curved spatial metric and shift be such at some event in that moment that they tend to make the curved cone violate the flat one a bit later.  By suitably reducing the lapse, one can lengthen the curved cone until it once again is safely inside the flat cone.  By so choosing the lapse at all times and places, one should be able to
satisfy the causality principle at every event, if no global difficulties arise.  Global difficulties are unlikely to arise, because the special relativistic approach excludes many nontrivial global spacetime features \emph{a priori}.  This plausibility argument need only work for sufficiently well-behaved solutions, for which it appears to be adequate.  Geroch and Ashtekar point out that some manifolds admitting curved metrics of Lorentzian signature do not admit flat ones \cite{AshtekarGeroch,GerochSpinor}, but fail to show that such manifolds must be regarded as physically admissible. 
Well-behaved solutions, at a minimum, have no closed causal curves, are orientable and time-orientable, are deformable to Minkowski spacetime, and have Lorentzian signature and a Cauchy surface.  

	This lapse-reducing plausibility argument especially ought to work for bounded matter distributions. Given that unbounded matter sources are standard in cosmology, one ought also to consider the homogeneous Robertson-Walker models.  For the Robertson-Walker cosmological spacetimes, the spatially flat case is permitted;  the negatively curved case most likely is permitted, with perhaps an inhomogeneous relation between the two metrics; while the positively curved case most likely is not.  (Positively curved solutions clearly are admissible if the requirement of homogeneity is relaxed, so the observation of energy density greater than the critical value would not pose any serious objection to this project.)  As we discuss elsewhere \cite{IARD2002}, fitting Robertson-Walker solutions into Minkowski spacetime implies that a fundamental observer's worldline will have finite length according to the observable curved metric $g_{\mu\nu}$ and infinite length according to the flat metric $\eta_{\mu\nu}$.  This finite $g$-length is not worrisome in the same way that the Big Bang singularity is in the geometrical approach, because of the intuition \cite{Wald} that field misbehavior `at infinity' is nonsingular \cite{IARD2002}. Let us borrow useful terminology from Gotay and Demaret \cite{GotayDemaretPRD,GotayDemaretNPB}: $\eta$-proper time acts like a ``fast'' time, because the classical singularity occurs in the infinite past or future, not in the finite past or future as for a ``slow'' time.

	Turning to the Schwarzschild solution, one might worry that a static version of the solution in the SRA would exclude even the region within the horizon and thus omit black holes \cite{RosenSchwarzschild,Bel,JNW,CooperstockJunevicus}.  However, no good reason to impose staticity exists, especially in the context of gravitational collapse.  Instead, using the  Loskutov-Vlasov-Petrov 
\cite{VlasovIncorrect,PetrovHarmonic,LoskutovSchwarzschild} \emph{stationary} harmonic Schwarzschild solution, and making a formal translation of the radial coordinate for the flat metric only, one can include the region between the horizon and the singularity.  These authors' form of the curved metric is

\begin{eqnarray}
ds^{2} = -\frac{r-M}{r+M}dt^{2} +  \frac{8M^{2} }{(r+M)^{2} }dt dr  \nonumber \\  + \left[ 1 + \frac{2M}{r + M}   +  \frac{4M^{2} }{(r + M)^{2} }      +  \frac{8 M^{3} }{ (r + M)^{3} }  \right] dr^{2} + (r+M)^{2}d\Omega^{2}. 
\end{eqnarray}
This curved metric, mated to a new flat metric $\eta_{\mu\nu}^{\prime}$ with line element 
\begin{equation} d\sigma^{\prime 2} = -dt^{2} + dr^{2} + (r+M)^{2}d\Omega^{2} \end{equation}
 with $r \geq -M,$ ensures that 
 $\eta$-causality is respected everywhere down to the true singularity at $r = -M,$ as one readily shows using the eigenvalue formalism below and making some plots using \emph{Mathematica}. An obvious change of coordinates $\rho = r + M$ would display both line elements using standard spherical polar coordinates and put the singularity at $\rho=0$.  Thus black holes are permitted in the special relativistic approach to Einstein's equations.  Petrov further finds \cite{PetrovHarmonic} that test particles reach the horizon in finite coordinate time $t$, but only hit  the singularity at $t= \infty$.  In Minkowski spacetime, there simply are no events with $t \geq \infty$, so it follows that, after crossing the horizon, an infalling test particle never hits the singularity, though it is always falling toward it and gets arbitrarily close to it.  This result is very satisfactory, because one need not ask what happens to the particle \emph{at} the singularity. The particle is not destroyed, only trapped.  Of course quantum gravity effects are likely to become important, but it is pleasant to find that the classical theory does not predict its own breakdown.  

	In discussing the Schwarzschild solution, we ought to recall \cite{SliBimGRG} why it is not representative of any deep worry for Lorentz-covariant approaches to Einstein's equations, \emph{pace}  R. Penrose \cite{Penrose} and (following him) J. Bi\v{c}\'{a}k \cite{Bicak}.  Penrose, recognizing 
a practical connection between scattering theory and the Lorentz-covariant approach 
to gravity,  poses a dilemma for the latter. 
 Using global techniques, he shows that either the curved null cone locally violates the flat one, or the scattering properties become inconvenient because the geodesics for the two metrics
 continue to diverge even far away from a localized source.  He concludes 
that a ``satisfactory'' relationship between the two null cones cannot be found.  Moreover, this situation is not specific to the Schwarzschild solution, but holds rather generally, given positive mass. Clearly the first horn of Penrose's dilemma would be fatal for any project that regards the flat metric as physically significant, rather than merely instrumentally useful in a field formulation of general relativity.  The second horn, however, does not constitute a fundamental problem, so we simply accept the second horn \cite{SliBimGRG}.  It is well known that long-range fields have inconvenient scattering properties \cite{Goldstein}.  The root of the divergence between the geodesics is merely the long-range $\frac{1}{r}$ character of the potential in the conformally invariant part of the curved metric, which one expects in a massless spin 2 theory.    

	By carefully considering which solutions (or pieces thereof) are physically necessary, plausibly one concludes that this use of the lapse (and perhaps the shift) to achieve null cone consistency suffices in general.  Some valid solutions of general relativity might lack analogs in the special relativistic approach.  For example, Ionescu has found it difficult to include an infinite plate source in Minkowski spacetime without violating the null cone structure of the background \cite{Ionescu}.  But even if an infinite plate source cannot be admitted, this is not a serious problem, because, as Griffiths points out in an electromagnetic context, such problems are artificial \cite{Griffiths} (p. 87).  Other solutions of Einstein's equations might not wholly fit into Minkowski spacetime, but a suitable piece of the solution will fit, as we saw for the Schwarzschild solution.  One wants solutions in the special relativistic approach to be extended as far as possible in length measured by the curved metric $g,$ but no further, so one takes the largest piece of the solution that will fit into Minkowski spacetime.  The requirement that the entirety of Minkowski spacetime (except perhaps for isolated singular points) be covered involves the metric and topological structures, not just the causal structure, of Minkowski spacetime.  Finally, S. Krasnikov has pointed out the need for a suitable energy condition to prevent tame spacetimes from spontaneously generating exotic features (see also (\cite{HawkingChronology}). 

	Likely this procedure of reducing the lapse to obtain the correct null cone relationship will force the lapse toward $0$ in some cases.  Then physical happenings (described intrinsically) are stretched out over more and more of Minkowski spacetime, perhaps to future or past infinity.  This procedure therefore bears a formal resemblance to the use of singularity-avoiding coordinates in
 numerical relativity \cite{Lehner}, in which the gauge (coordinate) freedom is used to exile singularities to infinite coordinate values.  However, the move here will prove to be gauge invariant, once a suitable redefinition of gauge transformations is made.  
	

\subsection{The Causality Principle and Loose Inequalities}

	As should be clear from the worries about conformally flat curved metrics, the desired relationship between the two null cones takes the form of some loose inequalities $a \leq b$.  Such relations have been called ``unilateral''
 \cite{OdenReddy} or ``one-sided'' constraints \cite{Lacomba},  typical examples being nonpenetration conditions.  Such constraints are more difficult to
 handle than the standard ``bilateral'' or ``two-sided'' constraint \emph{equations} that most treatments of constraints in physics discuss. 
Loose inequalities $a \leq b$  are also more difficult to handle than strict inequalities $a < b,$ such as the  positivity conditions in canonical general relativity \cite{GoldbergKlotz,IshamKakas,Kuchar92,Klauder2}, which require that the ``spatial'' metric have a positive determinant $h>0.$  One might
 eliminate the positivity conditions by a change of variables \cite{GoldbergKlotz} that satisfies the inequalities identically, such as an exponential function  $h = e^{y}.$  One could make such an exponential change of variables involving the eigenvalues from the formalism below.

\subsection{New Variables and the Segr\'{e} Classification of the Curved Metric with Respect to the Flat}

	On account of the many cross terms, the existing formalism for describing the relation between the two null cones  \cite{LogunovBasic,LogunovFund} has not been perfectly convenient.  
 One could achieve a slight savings by using the conformally invariant weight $-\frac{1}{2}$ densitized part of the metric $ g_{\mu\nu} (-g)^{-\frac{1}{4}}.$  Then only
nine numbers at each event are 
required (the determinant being $-1$), but that is still too many.    One would like to diagonalize
$g_{\mu\nu}$ and $\eta_{\mu\nu}$ simultaneously by solving the generalized eigenvalue problem 
\begin{equation}
 g_{\mu\nu} V^{\mu} = \Lambda \eta_{\mu\nu} V^{\mu},
\end{equation}
or perhaps the related problem using $ g_{\mu\nu} (-g)^{-\frac{1}{4}}$ and $ \eta_{\mu\nu} (-\eta)^{-\frac{1}{4}}.$  (The flat metric tensor $\eta_{\mu\nu}$ can be expressed in any coordinates, not only Cartesian coordinates.)   However, in general that is
impossible, because there is not a complete set
of eigenvectors, due to the indefinite (Lorentzian) nature of both tensors \cite{Eisenhart,HallArab,HallDiff,HallNegm,Hall5d,BonaCollMorales}.
There are four 
Segr\'{e} types for a real symmetric rank 2 tensor with respect to a Lorentzian metric,
the several types having 
different numbers and sorts of
eigenvectors \cite{HallArab,HallDiff,HallNegm,Hall5d}. 

 	To our knowledge, the only previous work to consider a generalized eigenvector decomposition of a curved Lorentzian metric with respect to a flat one was done by  I.
Goldman \cite{Goldman} in the context of Rosen's bimetric theory
of gravity, which does not use Einstein's field equations.
The lack of gauge freedom in Rosen's theory also ensured that the curved null cone was not
 subject to adjustment, unlike the situation in general relativity with a flat metric.  The linearized plane wave solution will appear in Rosen's theory, but there being no gauge freedom, one can find no gauge-related solution with the proper null cone relationship.  Thus Rosen's theory is not consistent with special relativity.  Any theory of gravity with an observable flat metric will almost certainly suffer the same fate.  Ironic as it may be, these theories, which wear their devotion to the background metric proudly in their field equations, in fact violate  special relativistic causality.  On the other hand, a theory based on Einstein's equations, in which the flat metric is shrouded behind the gravitational potential, can respect special relativistic causality, because Einstein's theory evidently has enough gauge freedom to secure null cone consistency.

	An eigenvalue-eigenvector decomposition for the \emph{spatial} metric was briefly contemplated by Klotz and Goldberg \cite{GoldbergKlotz}.  For space, 
as opposed to spacetime, one has a positive definite background (identity) matrix, so the usual theorems about eigenvectors and eigenvalues apply. But Klotz and Goldberg, who did not assume a background metric to exist,
 found little use for the eigenvector decomposition because of the nontensorial nature of the $3 \times 3$ identity matrix.  Such a decomposition, even given a background metric, is still somewhat complicated unless the ADM shift vanishes. 
 	
	Let us now proceed with the diagonalization project.  (A very brief assertion without proof of a few results from the eigenvector formalism appeared earlier \cite{NCTRA}.)  Given that a complete set of generalized eigenvectors might fail to exist, it is necessary to consider how many eigenvectors do exist and
under which conditions. This problem has been substantially addressed in a different context by G.
S. Hall and collaborators \cite{HallArab,HallDiff,HallNegm,Hall5d}, who were
interested in classifying the stress-energy  or Ricci tensors with respect to
the (curved) metric in (geometrical) general relativity. Such problems have in fact been studied over quite a long period of time \cite{BonaCollMorales} (and references therein), but we find the work of Hall \emph{et al.} to be especially convenient for our purposes.   
 There exist four cases, corresponding to the four
possible Segr\'{e} types (apart from degeneracies)  for the classified tensor.
The case $\{1,111\}$ has a complete set of eigenvectors ($1$ timelike, $3$
spacelike with respect to $\eta$), and is thus the most convenient case.  The
case  $\{211\}$ has two spacelike eigenvectors and one null eigenvector (with respect to $\eta$), whereas the  $\{31\}$ case has 
one spacelike eigenvector and one null one.  The last case, $\{z\,\bar{z}11\}$   has 2 spacelike eigenvectors with real eigenvalues and 2 complex eigenvalues.

	We now consider the conditions under which metrics of each of these Segr\'{e} classes obey $\eta$-causality.  To give a preview of our results, we state
that the $\{1,111\}$ and $\{211\}$ cases sometimes do obey it, although the
$\{211\}$  metrics appear to be dispensable, being borderline cases.  But no metric of type $\{31\}$ or
$\{z\,\bar{z}11\}$  obeys
 the causality principle, so these types can be
excluded from consideration.  

 	Hall \emph{et al.} introduce a real null tetrad of vectors $L^{\mu}, N^{\mu}, X^{\mu}, Y^{\mu}$ 
with vanishing inner products, apart from the relations
\begin{equation} \eta_{\mu\nu} L^{\mu} N^{\nu} =
\eta_{\mu\nu} X^{\mu} X^{\nu} = \eta_{\mu\nu} Y^{\mu} Y^{\nu} = 1.
\end{equation}
 Thus $L^{\mu}$ and $N^{\mu}$ are null, while
$X^{\mu}$ and $Y^{\mu}$ are spacelike. 
 (The signature is $-+++$.)  Expanding
an arbitrary vector $V^{\mu}$ as 
\begin{equation}
V^{\mu} = V^{L} L^{\mu} + V^{N} N^{\mu} +
V^{X} X^{\mu} + V^{Y} Y^{\mu}
\end{equation}
and taking the $\eta$-inner product with each vector
of the null tetrad gives \mbox{ $V^{L} = \eta_{\mu\nu} V^{\mu} N^{\nu} $,}  \mbox{ $V^{N} = \eta_{\mu\nu} V^{\mu} L^{\nu}$, }  \mbox{ $V^{X} = \eta_{\mu\nu} V^{\mu} X^{\nu} $,}
 and \mbox{ $V^{Y} = \eta_{\mu\nu} V^{\mu} Y^{\nu}$. } 
Thus, the Kronecker delta
tensor can be written as 
\begin{equation}
\delta^{\mu}_{\nu} = L^{\mu} N_{\nu} + L_{\nu}
N^{\mu} + X^{\mu} X_{\nu} + Y^{\mu} Y_{\nu},
\end{equation}
 indices being lowered here  using $\eta_{\mu\nu}$.  For some purposes it is also convenient to
define the timelike vector \mbox{ $T^{\mu} = \frac{L^{\mu} - N^{\mu} }{\sqrt{2} }$ } and  the spacelike vector \mbox{  $Z^{\mu} = \frac{ L^{\mu} + N^{\mu} }{\sqrt{2}}$. }

	We employ the results of Hall \emph{et al.} \cite{HallArab,HallDiff,HallNegm,Hall5d}, who find that 
the four possible  Segr\'{e} types (ignoring degeneracies) for a
(real) symmetric rank 2 tensor in a four-dimensional spacetime with a Lorentzian
metric can be written in the following ways, using a well-chosen
null tetrad.  The type $\{1,111\}$ can be written as
\begin{eqnarray}
g_{\mu\nu} = 2\rho_{0} L_{(\mu} N_{\nu)} + \rho_1 (L_{\mu} L_{\nu} + N_{\mu} N_{\nu} )
+ \rho_2 X_{\mu} X_{\nu} + \rho_3 Y_{\mu} Y_{\nu},
\end{eqnarray} or equivalently
\begin{eqnarray}
g_{\mu\nu} = -(\rho_{0} - \rho_{1} ) T_{\mu} T_{\nu} +  (\rho_{0} + \rho_{1} ) Z_{\mu} Z_{\nu} + 
\rho_2 X_{\mu} X_{\nu} + \rho_3 Y_{\mu} Y_{\nu}.
\end{eqnarray}
As usual, the parentheses around indices mean that the symmetric part should be taken \cite{Wald}.
The type $\{211\}$ can be written as 
\begin{eqnarray}
g_{\mu\nu} = 2\rho_{1} L_{(\mu} N_{\nu)} + \lambda L_{\mu} L_{\nu} 
+ \rho_2 X_{\mu} X_{\nu} + \rho_3 Y_{\mu} Y_{\nu}, 
\end{eqnarray} with $\lambda \neq 0$, the null eigenvector being $L^{\mu}$.
The type $\{31\}$ can be written as 
\begin{eqnarray}
g_{\mu\nu} = 2\rho_{1} L_{(\mu} N_{\nu)}  + 2 L_{(\mu} X_{\nu)}  
+ \rho_1 X_{\mu} X_{\nu} + \rho_2 Y_{\mu} Y_{\nu},  
\end{eqnarray} 
the null eigenvector again being $L^{\mu}$.
 The final type, $\{z\,\bar{z}11\}$, can be written as
\begin{eqnarray}
g_{\mu\nu} = 2\rho_{0} L_{(\mu} N_{\nu)} + \rho_1 (L_{\mu} L_{\nu} - N_{\mu} N_{\nu} )
+ \rho_2 X_{\mu} X_{\nu} + \rho_3 Y_{\mu} Y_{\nu},
\end{eqnarray} with $\rho_1 \neq 0$.
The requirements to be imposed upon the curved metric  for the moment are the following:  all $\eta$-null vectors must be $g$-null or $g$-spacelike, all $\eta$-spacelike eigenvectors must be $g$-spacelike, $g_{\mu\nu}$ must be 
Lorentzian (which amounts to having a negative determinant), and $g_{\mu\nu}$ must be connected to $\eta_{\mu\nu}$ by a succession of small changes which respect $\eta$-causality and the Lorentzian signature.  
It convenient to employ a slightly redundant form that admits all four types in order to treat them simultaneously.  Thus, we write
\begin{eqnarray}
g_{\mu\nu} = 2 A L_{(\mu} N_{\nu)} + B L_{\mu} L_{\nu} +  C N_{\mu} N_{\nu} 
+ D X_{\mu} X_{\nu} + E Y_{\mu} Y_{\nu} + 2 F  L_{(\mu} X_{\nu)}.
\end{eqnarray}  Using this form for $g_{\mu\nu}$, one
readily finds the squared length of a vector $V^{\mu}$ to be
\begin{eqnarray}
g_{\mu\nu} V^{\mu}V^{\nu} = 2 A V^{L} V^{N} + B (V^{N})^{2} + C (V^{L})^{2} + D (V^{X})^{2} 
+ E (V^{Y})^{2} \nonumber \\  + 2 F V^{X} V^{N}.
\end{eqnarray}
It is not clear \emph{a priori} how to express sufficient conditions for the causality principle 
in a convenient way.  
But it will turn out that the necessary conditions that we can readily impose are also sufficient.

\subsection{Necessary Conditions for Respecting the Flat Metric's Null Cone}

	The causality principle requires that the $\eta$-null vectors $L^{\mu}$ and $N^{\mu}$ be $g$-null or $g$-spacelike, 
so $B \geq 0, C \geq 0$.  These conditions already exclude the type   
$\{z\,\bar{z}11\}$, because the form above 
 requires that $B$ and $C$ differ in sign.  It must also be the case that the $\eta$-spacelike vectors 
$X^{\mu}$ and $Y^{\mu} $ are $g$-spacelike, so  $D>0$ and $E>0$.  

	Not merely  $L^{\mu}$ and $N^{\mu}$, but all $\eta$-null vectors must be $g$-null or $g$-spacelike. 
 This requirement quickly implies that $E \geq A$, and also requires that 
\begin{eqnarray}
 B (V^{N})^{2} + 2 F V^{X} V^{N} + (D-A) (V^{X})^{2} \geq 0.
\end{eqnarray} 
Here there are two cases to consider:  $F=1$ for type $\{31\},$ and $F=0$ for types  $\{1,111\}$ and $\{211\}.$  Let us consider $F=1.$  The $\{31\}$ has $B=0$, so the equation reduces to $2 F V^{X} V^{N} + (D-A) (V^{X})^{2} \geq 0, $ which implies that either $ V^{X} = 0$ or, failing that, $2 F V^{N} + (D-A) V^{X} \geq 0. $   Clearly one could also consider a null vector with the opposite value of $V^{X},$ 
yielding the inequality   $2 F V^{N} - (D-A) V^{X} \geq 0. $  Adding these two inequalities gives $4 V^{N} \geq 0,$ which simply cannot be made to hold for all values of $V^{N}.$  Thus, the $F=1$ case yields no $\eta$-causality-obeying curved metrics, and the   $\{31\}$ type is eliminated.  It remains to consider $F=0$ for the $\{1,111\}$ and $\{211\}$ types.  The resulting inequality is 
$B (V^{N})^{2}  + (D-A) (V^{X})^{2} \geq 0.$  Because $B \geq 0$ has already been imposed, it follows only that $D \geq A.$

	Let us summarize the results so far.  The inequalities  $B \geq 0$ and $C \geq 0$ have excluded the type $\{z\,\bar{z}11\}.$  We also have $D > 0,$ $D \geq A,$ $E > 0,$ $ E \geq A.$ Finally, $F=0$ excludes the type $\{31\},$ so only $\{1,111\}$ and $\{211\}$ remain. 

	We now impose the requirement of Lorentzian signature.  At a given event, one can find a  coordinate $x$ such that $ (\frac{\partial}{\partial x})^{\mu} = X^{\mu}$ and (flipping the sign of $Y^{\mu}$ if needed for the orientation)  a coordinate $y$ such that $ (\frac{\partial}{\partial y})^{\mu} = Y^{\mu};$ these two coordinates can be regarded as Cartesian at that event.  Then the null vectors $L^{\mu}$ and $N^{\mu}$ lie in the $t-z$ plane.  The curved metric has a block diagonal part in the $x-y$ plane with positive determinant, so imposing a Lorentzian signature means ensuring a negative determinant for the $2 \times 2$ $t-z$ part.  The vectors $L^{\mu}$ and $N^{\mu}$ in such a coordinate systems take the form $L^{\mu} = (L^{0}, 0,0,L^{3})$ and $N^{\mu} = (N^{0}, 0 , 0, N^{3}).$  Given
the Cartesian form $\eta_{\mu\nu} = diag(-1,1,1,1)$ and the nullity of these two vectors, it follows that $|L^{0}| = |L^{3}|$ and $|N^{0}| = |N^{3}|.$  Therefore the relevant parts of the curved metric can be written in such a coordinate basis as 
\begin{eqnarray*}
g_{00} = 2 A L^{0} N^{0} + B (L^{0})^{2} + C (N^{0})^{2},  \nonumber \\
g_{03}=g_{30}= -A(N^{0} L^{3} + L^{0} N^{3}) - B L^{0} L^{3} - C N^{0} N^{3}, \nonumber \\
g_{33} = 	2 A L^{3} N^{3} + B (L^{3})^{2} + C (N^{3})^{2}. \nonumber
\end{eqnarray*}  
Taking the determinant using \emph{Mathematica} and recalling that  $|L^{0}| = |L^{3}|$ and $|N^{0}| = |N^{3}|,$ one finds that the condition for a negative determinant is $2(A^{2} - BC) |L^{3}|^{2} |N^{3}|^{2} (sign(L^{0} L^{3} N^{0} N^{3}) -1) < 0.$  The linear independence of $L^{\mu}$ and $N^{\mu}$ implies that $sign(L^{0} L^{3} N^{0} N^{3}) = -1,$ so the determinant condition is $A^{2} -BC >0.$  Because $B$ and $C$ are both nonnegative, \mbox{ $A^{2} -BC >0$ } implies that $A \neq  0.$  But the requirement that the curved metric be smoothly deformable through a sequence of signature-preserving steps  means that the curved metric's value of $A$ cannot ``jump'' from one sign of $A$ to another, but must agree with the flat metric's positive sign.  It follows that $A > 0.$  

	We now summarize the necessary conditions imposed:
\begin{eqnarray}
A>0,  & A^{2}>BC, & B\geq 0, \nonumber \\ C \geq 0, & D \geq A, & E \geq A,  \nonumber \\ 
& F=0. 
\end{eqnarray}

\subsection{Sufficient Conditions for Respecting the Flat Metric's Null Cone}

	Thus far, it is not clear whether these necessary conditions are sufficient.  We now prove that they are.  It is helpful to consider the two types, $\{1,111\}$ and $\{211\}$, separately.  

	For the type $\{1,111\}$, the conditions on the coefficients $A,B,$ \emph{etc.} reduce to 
\begin{eqnarray}
A>0, & A>B, & B\geq 0, \nonumber \\ C = B, & D \geq A, & E \geq A.
\end{eqnarray}  For this form the following relations between variables hold:
\begin{eqnarray} A = \rho_{0}, & B = \rho_{1}, & D = \rho_{2}, \nonumber \\ 
& E = \rho_{3}.
\end{eqnarray}  It follows that this type can be expressed as 
\begin{eqnarray}
g_{\mu\nu} = -(A-B) T_{\mu} T_{\nu} +  (A + B) Z_{\mu} Z_{\nu} + 
D X_{\mu} X_{\nu} + E Y_{\mu} Y_{\nu}. 
\end{eqnarray}
Writing the eigenvalues for $T^{\mu},$ $X^{\mu},$ $Y^{\mu},$ and $Z^{\mu}$ as $D^{0}_{0},$ $D^{1}_{1},$ $D^{2}_{2},$ and $D^{3}_{3}, $ respectively, one has 
\begin{eqnarray}
D^{0}_{0} = A-B, & D^{1}_{1} = A + B,  & D^{2}_{2} = D, \nonumber \\
& D^{3}_{3} = E.
\end{eqnarray}
One sees that the inequalities imply that the eigenvalue for the timelike eigenvector $T^{\mu}$  (briefly, the ``timelike eigenvalue'') is no larger than any of the spacelike eigenvalues:
\begin{eqnarray}
D^{0}_{0} \leq D^{1}_{1}, & D^{0}_{0} \leq D^{2}_{2}, &  D^{0}_{0} \leq D^{3}_{3}, 
\end{eqnarray}
and that all the (generalized) eigenvalues are positive.
Let us now see that these conditions are sufficient.  Writing an arbitrary vector $V^{\mu}$ as
\begin{equation}
V^{\mu} = V^{T} T^{\mu} + V^{X} X^{\mu} + V^{Y} Y^{\mu} + V^{Z} Z^{\mu},
\end{equation}
 one sees that its $\eta$-length (squared) is 
\begin{equation}
\eta_{\mu\nu} V^{\mu}V^{\nu} = -(V^{T})^{2} + (V^{X})^{2} + (V^{Y})^{2} + (V^{Z})^{2}.
\end{equation}  Clearly this length
 is never more positive than 
\begin{equation}
\frac{1}{D^{0}_{0} }  g_{\mu\nu} V^{\mu}V^{\nu} = -(V^{T})^{2} + \frac{ D^{1}_{1} }{D^{0}_{0} } (V^{X})^{2} + \frac{ D^{2}_{2} }{D^{0}_{0} }  (V^{Y})^{2} + \frac{ D^{3}_{3} }{D^{0}_{0} } (V^{Z})^{2}, 
\end{equation} so the necessary conditions are indeed sufficient for type    $\{1,111\}.$

	For the type $\{211\}$, the conditions on the coefficients $A,B,$ \emph{etc.} reduce to 
\begin{eqnarray}
A>0, & B>0, & C = 0, \nonumber \\ D \geq A, & E \geq A, & F=0.
\end{eqnarray}
One can write the curved metric in terms of $T^{\mu}$, $Z^{\mu}$, $X^{\mu}$, and $Y^{\mu}$, 
though $T^{\mu}$ and $Z^{\mu}$ are not eigenvectors.  One then has
\begin{eqnarray}
g_{\mu\nu} = \left(-A + \frac{B}{2} \right) T_{\mu} T_{\nu} +  \left(A + \frac{B}{2} \right) Z_{\mu} Z_{\nu}  + B Z_{(\mu} T_{\nu)}   + \nonumber \\
  D X_{\mu} X_{\nu} + E Y_{\mu} Y_{\nu}. \end{eqnarray}
Writing an arbitrary $\eta$-spacelike vector field $V^{\mu}$ as 
\begin{equation}
V^{\mu} = G T^{\mu} + H Z^{\mu} + I X^{\mu} + J Y^{\mu}, \end{equation}
 with $H^{2} + I^{2} + J^{2} > G^{2}$, one readily finds the form of
$g_{\mu\nu} V^{\mu} V^{\nu} $.  Employing the relevant inequalities and shuffling coefficients, one
obtains the manifestly positive result 
\begin{equation}
g_{\mu\nu} V^{\mu} V^{\nu} = A(H^{2} + I^{2} + J^{2} -
G^{2}) + \frac{1}{2}B(G-H)^{2} + (D-A)I^{2} + (E-A) J^{2}. \end{equation}
  This positivity
result says that all $\eta$-spacelike vectors are $g$-spacelike.  Earlier the
requirement that all $\eta$-null vectors be $g$-null or $g$-spacelike was
imposed.  These two conditions together comprise the causality principle, so
we have obtained sufficient conditions for the $\{211\}$ type also.  

	The $\{211\}$ type, which has with one null and two spacelike eigenvectors,  is a borderline case in which the curved metric's null cone is tangent to the flat metric's cone along a single direction \cite{ChurchillVector}. 
 Clearly such borderline cases of $\{211\}$ metrics obeying the causality principle form in some sense  a measure $0$ set of all causality principle-satisfying metrics. 
 Given that they are so scarce, one might consider neglecting them.  Furthermore, they are arbitrarily close to violating the causality principle.  We recall the 
 criterion of stable causality in geometrical general relativity \cite{Wald} (where the issue is closed timelike curves, without regard to any flat metric's null cone), which frowns upon metrics which satisfy causality, but would fail to do so if
 perturbed by an arbitrarily small amount.  One could imagine  that quantum fluctuations might push such a marginal metric over the edge, and thus prefers to exclude such metrics as unphysical.
  By analogy, one might impose stable $\eta$-causality, which excludes curved metrics that are arbitrarily close to violating the flat null cone's 
notion of causality, though we saw that such a condition would exclude conformally flat metrics, also.  
Another reason for neglecting type $\{211\}$ metrics is that they are both  technically inconvenient and physically unnecessary.  Because $\eta$-causality-respecting $\{211\}$ metrics are arbitrarily close to $\{1,111\}$ metrics, one could merely make a small naive gauge 
transformation
to shrink the lapse a bit more and obtain a $\{1,111\}$ metric instead.    Thus, every curved metric that respects $\eta$-causality either is of type $\{1,111\},$ or is arbitrarily close to being of type  $\{1,111\}$  and deformable thereto by a small naive gauge transformation reducing the lapse.  

	It follows that there is no loss of generality in restricting the configuration space to type  $\{1,111\}$  curved metrics,   for which the two metrics are simultaneously diagonalizable.  As a result, there exists a close relationship between 
the traditional orthonormal tetrad formalism and this eigenvector decomposition. 
 In particular, one can build a $g$-orthonormal tetrad field $e^{\mu}_{A}$ simply by choosing the normalization of the eigenvectors. This choice fixes  the local Lorentz freedom of the tetrad (except
 when eigenvalues are degenerate) in terms of the flat metric tensor. 

 Rewriting the generalized eigenvector equation for the case in which a complete set exists, one
 can write 
\begin{equation}
g_{\mu\nu} e^{\mu}_{A} = \eta_{\mu\nu} e^{\mu}_{B} D^{B}_{A},
\end{equation} with the four eigenvalues being the elements of the diagonal matrix $D^{A}_{B}$.  It is sometimes convenient to raise or lower the indices of this matrix using the matrix $\eta_{AB} = diag(-1,1,1,1)$.  
The tetrad field has $\{  e^{\mu}_{A} \}$ has  
inverse $\{  f_{\mu}^{A} \}$.    We recall the standard relations $g_{\mu\nu} = f_{\mu}^{A} \eta_{AB} f_{\nu}^{B} $ and   $g_{\mu\nu} e^{\mu}_{A} e^{\nu}_{B} = \eta_{AB}.  $
 It is not difficult to show the how the tetrad lengths are related to the eigenvalues:  $\eta_{\mu\nu} e^{\mu}_{A} e^{\nu}_{B} = D^{-1}_{AB},$ and equivalently, $\eta^{\mu\nu} f_{\mu}^{A} f_{\nu}^{B} = D^{AB}.$  It follows that 
 $f_{\nu}^{A}  = \eta_{\nu\alpha}  e^{\alpha}_{B} D^{AB}$, which says that a given leg of the cotetrad $f_{\nu}^{A}$ can be expressed solely in terms of the corresponding leg of the tetrad $e^{\mu}_{A}$, through a stretching, an index lowering, and possibly a sign change, without reference to the other legs.
Simultaneous diagonalization, of course,  implies that the tetrad vectors are orthogonal to each other with respect to both metrics.  

 \subsection{Finite Gauge Transformations and a Tetrad}

	While linearized gauge transformations are well known in field formulations of general relativity, the form of a \emph{finite} gauge transformation is less well known.  It has been shown by L. P. Grishchuk, A. N. Petrov, and A. D. Popova \cite{Grishchuk}  to have the form 
\begin{eqnarray}
	{{\mathfrak g}^{\sigma\rho}} \rightarrow e^{\pounds_{\xi}}   {\mathfrak g}^{\sigma\rho},
u \rightarrow e^{\pounds_{\xi}}   u,
	\eta_{\mu\nu} \rightarrow \eta_{\mu\nu}
\end{eqnarray}
in terms of the convenient variable ${\mathfrak g}^{\sigma\rho} = \sqrt{-g}  g^{\sigma\rho},$ the flat metric tensor, and matter fields $u$ described by some tensors (or tensor densities) with indices suppressed.  They made use of a first-order action.  Using a second-order form of the action, we will reconfirm that this transformation indeed changes the action merely be a boundary term.  We will also derive convenient formulas involving different sets of variables, so that one is not required to use ${\mathfrak g}^{\sigma\rho}$.  In addition, we will introduce the analogous formula for finite gauge transformations of tetrad fields.  

	J. L. Friedman (with D. M. Witt) has kindly pointed out that this exponentiated Lie derivative form is less general than one might expect intuitively.  In particular, there are diffeomorphisms near the identity that are not in the image of the exponential map \cite{Milnor,Freifeld}.  However, the counterexample given \cite{Milnor} involves closed spatial loops, whereas our need for gauge transformations involves timelike curves.  Probably we can simply confine our attention to those diffeomorphisms for which the exponential Lie formula holds in making gauge transformations in the SRA.  In that case, the exponential Lie formula holds by construction, so this qualification does not seem crucial for our purposes.

	We recall the bimetric form of the action above for a generally covariant theory \cite{SliBimGRG}, with the metric here expressed in terms of the weight 1 inverse metric density: 
\begin{equation}
S = S_{1} [{\mathfrak g}^{\mu\nu}, u] + \frac{1}{2} \int d^{4}x R_{\mu\nu\rho\sigma} (\eta)
{\mathcal{M}} ^{\mu\nu\rho\sigma} + 2 b \int d^{4}x \sqrt{-\eta}  + \int d^{4}x 
\alpha^{\mu},_{\mu}.
\end{equation}
${\mathcal{M}} ^{\mu\nu\rho\sigma} $ is an arbitrary weight 1 tensor density with the symmetries of the Riemann tensor, depending only on ${\mathfrak g}^{\sigma\rho},$ $u$, and $\eta_{\mu\nu}.$ 
Clearly the terms other than $S_{1}$ change at most by a boundary term, so our attention turns to  $S_{1} = \int d^{4}x {\mathcal{L}} _{1} $. 
 The important term $ {\mathcal{L}} _{1}$ in the Lagrangian density is 
just the sum of terms which are products of 
${\mathfrak g}^{\mu\nu}$, $u$, and their partial derivatives.

We now derive a useful formula.  Writing out $e^{\pounds_{\xi}}   A$  as a series 
\begin{equation}
e^{\pounds_{\xi}}   A = \sum_{i = 0}^{\infty} \frac{1}{i!} \pounds_{\xi}^{i} A \end{equation}
for some tensor density $A$, and a similar series for another tensor density $B$,  will put us in a position to derive a useful `product' rule for the exponential of Lie differentiation. (The index structures and density weights are arbitrary.) Multiplying the series and using the Cauchy product formula \cite{ChurchillComplex}
\begin{equation}
\sum_{i = 0}^{\infty} a_{i} z^{i} \sum_{j = 0}^{\infty} b_{j} z^{j} = 
\sum_{n = 0}^{\infty}   \sum_{k = 0}^{n} a_{k} b_{n-k} z^{k} 
\end{equation} and the $n$-fold iterated Leibniz rule \cite{ChurchillComplex}
\begin{equation}
[fg]^{(n)} =    \sum_{k = 0}^{n} \frac{n!}{k!(n-k)!} f^{(k)} g^{(n-k)},
\end{equation} 
one recognizes the result as the series expansion of $e^{\pounds_{\xi}}  (AB)$, so one has the pleasant result 
\begin{equation}
(e^{\pounds_{\xi}} A)   (e^{\pounds_{\xi}} B) = e^{\pounds_{\xi}} (AB)
\end{equation}

 Using the fact that partial differentiation commutes with Lie differentiation, 
 ones sees that replacing
 ${\mathfrak g}^{\mu\nu}$ by $e^{\pounds_{\xi}} {\mathfrak g}^{\mu\nu}$ and $u$ with  $e^{\pounds_{\xi}} u$ in ${\mathcal{L}} _{1}$ will give $e^{\pounds_{\xi}} {\mathcal{L}}_{1}$.  Thus, the change in $ {\mathcal{L}} _{1}$  is $\delta {\mathcal{L}} _{1} = (e^{\pounds_{\xi}} -1) {\mathcal{L}} _{1}$,  which is the Lie derivative of a scalar density of weight 1.  Recalling \cite{Israel} that the Lie derivative  a weight 1 scalar density $\phi$ is  $\pounds_{\xi} \phi = ( \xi^{\mu} \phi ),_{\mu}$, one sees that $\delta {\mathcal{L}} _{1} $ is just a coordinate divergence, as desired.  

	In view of the matrix relationships among the various metric quantities, one has by definition that $( {\mathfrak g}^{\mu\nu} + \delta {\mathfrak g}^{\mu\nu}) ( {\mathfrak g}_{\rho\nu} + \delta {\mathfrak g}_{\rho\nu})  = \delta^{\mu}_{\rho}$ and various other relations.  In that way, one can derive the form of 
 $\delta {\mathfrak g}_{\rho\nu} $, 
  $\delta g$,  $ \delta g_{\rho\nu} $, and the like.  Let us show this fact explicitly for $ g$,   using ${\mathfrak g}^{\sigma\rho} + \delta {\mathfrak g}^{\sigma\rho} =   e^{\pounds_{\xi}}   {\mathfrak g}^{\sigma\rho}.$
The determinant is given by $| {\mathfrak g}^{\sigma\rho}| = [\alpha\mu\nu\rho] \delta^{0}_{\beta} \delta^{1}_{\chi} \delta^{2}_{\psi} \delta^{3}_{\phi} {\mathfrak g}^{\alpha\beta} {\mathfrak g}^{\mu\chi} {\mathfrak g}^{\nu\psi} {\mathfrak g}^{\rho\phi}, $ where $[\alpha\mu\nu\rho] $ is the totally antisymmetric symbol with $[0123] =1.$   Because this form for the determinant holds in any coordinate system, $ [\alpha\mu\nu\rho] \delta^{0}_{\beta} \delta^{1}_{\chi} \delta^{2}_{\psi} \delta^{3}_{\phi} $ is a \emph{scalar} (and also a constant), so  $    e^{\pounds_{\xi}}    ( [\alpha\mu\nu\rho] \delta^{0}_{\beta} \delta^{1}_{\chi} \delta^{2}_{\psi} \delta^{3}_{\phi}  )       = [\alpha\mu\nu\rho] \delta^{0}_{\beta} \delta^{1}_{\chi} \delta^{2}_{\psi} \delta^{3}_{\phi} .$
We therefore have 
\begin{eqnarray}
|   e^{\pounds_{\xi}}        {\mathfrak g}^{\sigma\rho}| = [\alpha\mu\nu\rho] \delta^{0}_{\beta} \delta^{1}_{\chi} \delta^{2}_{\psi} \delta^{3}_{\phi} ({ e^{\pounds_{\xi}} \mathfrak g}^{\alpha\beta}) ( { e^{\pounds_{\xi}}   \mathfrak g}^{\mu\chi}) ( { e^{\pounds_{\xi}}  \mathfrak g}^{\nu\psi}) ( {  e^{\pounds_{\xi}}      \mathfrak g}^{\rho\phi}) \nonumber \\
=  e^{\pounds_{\xi}}      |     {\mathfrak g}^{\sigma\rho}| .
\end{eqnarray}
Using $ {\mathfrak g}^{\sigma\rho} =  g^{\sigma\rho} \sqrt{-g},$ one recalls that $| {\mathfrak g}^{\sigma\rho} | =     | g_{\sigma\rho} | , $
so
\begin{equation}
g + \delta g = e^{\pounds_{\xi}}      g.
\end{equation}
The relation $ -g - \delta g = (\sqrt{-g} + \delta \sqrt{-g} )^{2}$ defines $\delta \sqrt{-g}, $ so one quickly also finds that  
\begin{equation}
\sqrt{-g} + \delta \sqrt{-g} =  e^{\pounds_{\xi}}      \sqrt{-g} ,
\end{equation}
with which one readily finds the result for $ g^{\sigma\rho} $ and so on.  
 Again the transformed field is just the exponentiated Lie derivative of the original.  One therefore can readily use variables other than the densitized inverse curved metric ${\mathfrak g}^{\mu\nu}$.  

	Grishchuk, Petrov, and Popova  
have exhibited a straightforward and attractive relationship between finite gauge transformations (with the exponentiated Lie differentiation) and the tensor transformation law \cite{GrishchukPetrov,PopovaPetrov}.  Evidently the former is fundamental, the latter derived. 
 One can define a vector field $\xi^{\alpha}$ using the fact that under a gauge transformation, ${\mathfrak{g}}^{\mu\nu}$ changes in accord with the tensor transformation law, while the flat metric stays fixed.  Let us follow them and define $\xi^{\alpha}$  in terms of the finite coordinate transformation
\begin{equation}
 x^{\prime \alpha} = e^{  \xi^{\mu} \frac{\partial}{\partial x^{\mu} } } x^{\alpha}.
\end{equation}
Then the tensor transformation law gives the exponential Lie formula.  The tensor transformation law being easier to use in some respects, this connection is useful.

	It appears that finite gauge transformations for an orthonormal tetrad have never been studied before, so let us do so.  If one imposes no requirements on the tetrad other than that it be orthonormal, then the formula is nonunique in the local Lorentz transformation matrix field $F^{C}_{A}$.  The desired form turns out to be 
\begin{equation}
e^{\mu}_{A} + \delta e^{\mu}_{A} = e^{\pounds_{\xi} } (e^{F})^{C}_{A} e^{\mu}_{C}.
\end{equation}
The Lie differentiation in the first factor acts on everything to its right.  Thus one performs a finite local Lorentz transformation, and then acts with the Lie differentiation.
$F$ is a matrix field which, when an index is moved using $\eta_{AB} = diag(-1,1,1,1)$  or $\eta^{AB},$ is antisymmetric: $F_{A}^{C}= -\eta_{AE} F^{E}_{B}\eta^{BC}.$ 
One can show that the above formula preserves both the completeness
 relation to the inverse metric 
$g^{\mu\nu} = e^{\mu}_{A} \eta^{AB} e^{\nu}_{B}$ and the orthonormality relation $g_{\mu\nu} e^{\mu}_{A} e^{\nu}_{B} = \eta_{AB}.$ Let us now verify the completeness relation by showing that this relation with the gauge-transformed tetrad yields the gauge-transformed curved metric, using $(e^{\pounds_{\xi}} A)   (e^{\pounds_{\xi}} B) = e^{\pounds_{\xi}} (AB).$ One has by definition of a variation $\Delta$ induced by this tetrad transformation, \begin{eqnarray}
g^{\mu\nu} + \Delta g^{\mu\nu} = (e^{\mu}_{A} + \delta e^{\mu}_{A})  \eta^{AB} (e^{\nu}_{B} + \delta e^{\nu}_{B}) 
\nonumber \\
 = [e^{\pounds_{\xi} } (e^{ F} )^{C}_{A} e^{\mu}_{C}]	 	\eta^{AB}		 e^{\pounds_{\xi} } (e^{ F})^{E}_{B} e^{\nu}_{E}  \nonumber \\
= e^{\pounds_{\xi} } [(e^{ F})^{C}_{A} e^{\mu}_{C}	 	\eta^{AB}		  (e^{ F})^{E}_{B} e^{\nu}_{E}  ].
\end{eqnarray}
Acting with $e^{-\pounds_{\xi} } $ gives
\begin{eqnarray}
 e^{-\pounds_{\xi} }    (g^{\mu\nu} + \Delta g^{\mu\nu} ) = (e^{ F })^{C}_{A} e^{\mu}_{C}	 	\eta^{AB}		  (e^{ F })^{E}_{B} e^{\nu}_{E}.
\end{eqnarray}
  We shall use the  near-antisymmetry of $F$:  $F_{E}^{C} = - \eta_{EJ} F^{J}_{B} \eta^{BC}.$  One then has
\begin{eqnarray}
 e^{-\pounds_{\xi} }    (g^{\mu\nu} + \Delta g^{\mu\nu} )  =   e^{\mu}_{C} (e^{F})^{C}_{A} \eta^{AB} (e^{F})^{E}_{B}    e^{\nu}_{E}  \nonumber \\
=   e^{\mu}_{C} (e^{F})^{C}_{A} \eta^{AB} (I^{E}_{B}  + F^{E}_{B}  + F^{E}_{J} F^{J}_{B} + \ldots  )  e^{\nu}_{E} , 
\end{eqnarray}
where the one factor is expanded as a series.  Continuing by moving the Lorentz matrix $\eta_{AB}$ and its inverse into strategic locations gives 
\begin{eqnarray}   e^{\mu}_{C} (e^{F})^{C}_{A}  (I^{A}_{P} + \eta^{AB} F^{E}_{B} \eta_{EP} +  \eta^{AB}  F^{J}_{B} \eta_{JK} \eta^{KL} F^{E}_{L} \eta_{EP} + \ldots ) e^{P\nu} \nonumber \\
=  
 e^{\mu}_{C} (e^{F})^{C}_{A}  (I^{A}_{J} - F^{A}_{J}  +   F^{A}_{K} F^{K}_{J} - \ldots ) e^{J\nu}, 
\end{eqnarray}
where the near-antisymmetry of $F$ has been employed.
Reverting to the exponential form gives
\begin{eqnarray} 
 e^{\mu}_{C} (e^{F})^{C}_{A} (e^{-F})^{A}_{J} e^{J\nu} 
=  
 e^{\mu}_{C} I^{C}_{E} e^{E\nu}  
= g^{\mu\nu},
\end{eqnarray}
leading to the expected conclusion  $g^{\mu\nu} + \Delta g^{\mu\nu}  =   g^{\mu\nu} + \delta g^{\mu\nu}  = e^{\pounds_{\xi} }    g^{\mu\nu}.$  Thus, completeness holds, and the tetrad-induced variation $\Delta$ of the inverse curved metric agrees with the gauge transformation variation $\delta.$  
By similar maneuvers, one establishes the orthonormality relation for this tetrad variation:
\begin{eqnarray} 
 (  e^{\pounds_{\xi} }    g_{\mu\nu} )  (e^{\mu}_{A} + \delta e^{\mu}_{A}) (e^{\nu}_{B} + \delta e^{\nu}_{B}) =  \eta^{AB}.
\end{eqnarray}
Finally, the inverse tetrad transforms as 
\begin{equation}
f^{A}_{\mu} + \delta f^{A}_{\mu} = e^{\pounds_{\xi} } (e^{  -F })^{A}_{C} f^{C}_{\mu},
\end{equation}
with a negative sign applied to $F^{A}_{C}$.

 \subsection{Gauge Transformations Form a Groupoid}

	If one is not interested in taking $\eta$-causality seriously, then any suitably smooth vector field, perhaps subject to some boundary conditions, will generate a gauge transformation.    This is the notion that Grishchuk, Petrov, and Popova have employed, and that on occasion we have used above as a ``naive'' gauge transformation.  With suitable technical assumptions, these transformations should form a group.  However, in the SRA, respecting $\eta$-causality--indeed, preferably stable $\eta$-causality--is essential.  This fact entails that only a subset of all vector fields generates gauge transformations in the SRA. 

	Let us be more precise in defining gauge transformations in the SRA, requiring stable $\eta$-causality. A gauge transformation in the SRA is a mathematical transformation generated by a vector field in the form 
\[
g_{\mu \nu }\rightarrow e^{\pounds _{\xi}}g_{\mu\nu},
\eta _{\mu \nu}\rightarrow \eta _{\mu\nu},
u\rightarrow e^{\pounds _{\xi }}u, 
\]
but we now introduce the requirement that both the original and the transformed curved metrics 
respect stable $\eta$-causality.  It is evident that a vector field that generates a gauge transformation given one curved metric and a flat metric, might not generate a gauge transformation given another curved metric (and the same flat metric), because in the second case, the transformation might move the curved metric out of the  $\eta$-causality-respecting configuration space, which
 is only a subset of the naive configuration space. (One might also need to impose boundary conditions on the generating vector field to ensure maintenance of boundary conditions on the curved metric and matter fields.)

	It follows that one cannot identify gauge transformations with generating vector fields alone.  Rather, one must also specify the curved metric prior to the transformation.  For thoroughness, one can also use the flat metric as a label, to ensure that the trivial coordinate freedom is not confused with the physically significant gauge freedom.  Let us therefore  write a gauge transformation as an ordered triple involving a vector field, a flat metric tensor field, and a curved metric tensor field:
\begin{equation}
( e^{\pounds_{\xi} }, \eta_{\mu\nu}, g_{\mu\nu}),
\end{equation}
where both $g_{\mu\nu}$ and $e^{\pounds_{\xi} } g_{\mu\nu}$ satisfy stable causality with respect  to $\eta_{\mu\nu}.$ The former restriction limits the configuration space for the curved metric, whereas the latter restricts the vector field. (At this point we drop the indices  for brevity.)  The non-Abelian nature of these transformations makes it advisable to use not $\xi$, but the operator $e^{\pounds_{\xi}}$, to label the transformations, because then the noncommutativity of two transformations is manifest.   

	One wants to compose two gauge transformations to get a third gauge transformation.  At this point, the fact that a gauge transformation is not labelled merely by the vector field, but also by the curved and flat metrics, has important consequences.  Clearly the two gauge transformations to be composed must have the second one start with the curved metric with which the first one stops. 
 We also want the flat metrics to agree.  Thus, the `group' multiplication operation is defined only in certain cases, meaning the gauge transformations in the SRA \emph{do not form a group}, despite the inheritance of the mathematical form of exponentiating the Lie differentiation operator from the field formulation's gauge transformation. 
 Two gauge transformations $(e^{\pounds_{\psi} }, \eta_{2}, g_{2})$ and $(e^{\pounds_{\xi} }, \eta_{1}, g_{1})$ can be composed to give a new gauge transformation
\begin{equation}
(e^{\pounds_{\psi} }, \eta_{2}, g_{2}) \circ (e^{\pounds_{\xi} }, \eta_{1}, g_{1})  =  ( e^{\pounds_{\psi} } e^{\pounds_{\xi} },  \eta_{1}, g_{1})  
\end{equation}
if and only if $g_{2} = e^{\pounds_{\xi} } g_{1}$ and $\eta_{2} = \eta_{1}.$ 
  The left inverse of $(e^{\pounds_{\xi} }, \eta_{1},  g_{1} )$ is    $(e^{\pounds_{-\xi} }, \eta_{1}, e^{\pounds_{\xi} } g_{1} ),$ yielding
   \begin{equation}
(e^{\pounds_{-\xi} }, \eta_{1}, e^{\pounds_{\xi} } g_{1} ) \circ   (e^{\pounds_{\xi} }, \eta_{1},  g_{1} ) = ( 1, \eta_{1}, g_{1}),
\end{equation}
 an identity transformation.  The right inverse is  also $(e^{\pounds_{-\xi} }, \eta_{1}, e^{\pounds_{\xi} } g_{1} ),$ yielding 
\begin{equation}
 (e^{\pounds_{\xi} }, \eta_{1},  g_{1} ) \circ (e^{\pounds_{-\xi} }, \eta_{1}, e^{\pounds_{\xi} } g_{1} )  = (1, \eta_{1}, e^{\pounds_{\xi} }  g_{1} ),
\end{equation}
 which is also an identity  transformation.  Gauge transformations in the SRA do not form a group, because multiplication between some elements simply is not defined.  

	While the lack of a group structure is perhaps unfamiliar, there is a mathematical structure that precisely corresponds to what the physics of the SRA dictates.   According to A. Ramsay, ``[a] groupoid is, roughly speaking, a set with a not everywhere defined binary operation, which would be a group if the operation were defined everywhere.'' \cite{Ramsay} (pp. 254, 255)  One need not rest with informal descriptions, because one can easily show that SRA gauge transformations in fact satisfy the axioms required of a groupoid, as defined by P. Hahn \cite{Hahn} and J. Renault \cite{Renault}.  Though groupoids are increasingly coming to the attention of physicists, they are still sufficiently obscure that we reproduce the definition of Hahn and Renault here for convenience.  According to them \cite{Hahn,Renault}, a groupoid is a set $G$ endowed with a product map $(x,y)\rightarrow
xy:G^{2}\rightarrow G$, where $G^{2}$ is a subset of $G\times G$ called the
set of composable [ordered] pairs, and an inverse map $x\rightarrow
x^{-1}:G\rightarrow G$ such that the following relations are satisfied:
\begin{enumerate}
\item  $(x^{-1})^{-1}=x$,

\item  if $(x,y)$ and $(y,z)$ are elements of $G^{2}$, then $(xy,z)$ and $(x,yz)$ are elements of $G^{2}$ and $(xy)z=x(yz)$,

\item  $(x^{-1},x)\in G^{2}$, and if $(x,y)\in G^{2}$, then $x^{-1}(xy)=y$,

\item  $(x,x^{-1})\in G^{2}$, and if $(z,x)\in G^{2}$, then $(zx)x^{-1}=z$.
\end{enumerate}
One readily interprets this definition as implying that every SRA gauge transformation has an inverse, and that multiplication is associative whenever it is defined.  

	Some authors \cite{Grishchuk90,Petrov} have previously denied that the flat metric's null cone can have any physical significance, in part  because the relation between the two metrics' null cones is gauge variant.  If one tacitly assumes that gauge transformations must form a group, then that objection holds, but  insistence on a group structure seems unwarranted.  In accord with the demands of the SRA,  we define gauge transformations such that they respect the causal structure of the flat metric, and find that gauge transformations form a groupoid.   Thus this objection to ascribing physical significance to the flat metric's null cone is removed.  It would be interesting to describe the null cone relationship using the language of ``causal relationship'' of A. Garcia-Parrado and J. M. M. Senovilla \cite{SenovillaCausal} (and works cited therein).

\section{Equal Time Commutation Relations for Quantum Gravity}

	The use of a flat background metric in gravitation might suggest a connection to the old covariant perturbation program of quantum gravity.  In the modern canonical quantization program, ``background-dependent'' and ``perturbative'' are basically synonymous.  But in fact nothing about the SRA is inherently perturbative---nowhere is an expansion in powers of $\sqrt{G}$ or the like  made.  Moreover, expanding the gravitational potential in terms of plane waves (perhaps in the tensorial DeDonder gauge) and quantizing the Fourier coefficients in fact is very \emph{unnatural} in the SRA, because these plane waves (at least in the tensorial DeDonder gauge) individually violate $\eta$-causality.  It is far from  clear how to build an $\eta$-causal general solution out of pieces that are all $\eta$-acausal.  Thus a nonperturbative quantization of the SRA is not only permissible, but preferred over a perturbative one.

	The perturbative covariant quantization program famously appears to be nonrenormalizable, even  with the addition of carefully chosen matter fields in the later supergravity era \cite{IshamPrima}, so what can a flat background metric still contribute to 
quantum gravity?  Here is one answer: it can give a well-defined notion of causality.  Isham writes of the null cone consistency issue in the covariant perturbation program:  ``This very non-trivial problem is one of the reasons why the canonical approach to quantum 
gravity has been so popular.'' \cite{IshamPrima} (p. 12)  And again, ``One of the main aspirations of the canonical approach to quantum gravity has always been to build a formalism with no background spatial, or spacetime, metric.'' \cite{IshamPrima} (p. 18) The use of a flat background in canonical 
gravity indeed seems to be almost unknown, apart from a few works that do not consider the flat metric's null cone \cite{GrishchukPetrov,Solovev}.  However, it would be a mistake to conclude that the canonical formalism is immune to similar causality worries.  Isham continues:  
\begin{quote}
	However, a causal problem arises here [in the canonical approach] too.  For example, in the Wheeler-DeWitt approach, the configuration variable of the system is the Riemannian metric $q_{ab}(x)$ on a three-manifold $\Sigma,$ and the canonical commutation relations invariably include the set 
\begin{equation}
[\hat{q}_{ab}(x), \hat{q}_{cd}(x^{\prime} )] = 0
\end{equation}
for all points $x$ and $x^{\prime} $ in $\Sigma.$  In normal canonical quantum field theory such a relation arises because $\Sigma$ is a space-like subset of spacetime, and hence the fields at $x$ and $x \prime$ should be simultaneously measurable.  But how can such a relation be justified in a theory that has no fixed causal structure?  The problem is rarely mentioned but it means that, in this respect, the canonical approach to quantum gravity is no better than the covariant one.  It is another aspect of the `problem of time' \ldots. \cite{IshamPrima} (p. 12)
\end{quote}	
Evidently introducing a flat metric can help:
\begin{quote} The background metric $\eta$ provides a fixed causal structure with the usual family of Lorentzian inertial frames.  Thus, at this level, there is no problem of time.  The causal structure also allows a notion of microcausality, thereby permitting a conventional type of relativistic quantum field theory \ldots 
	It is clear that many of the \emph{prima facie} issues discussed earlier are resolved in an approach of this type by virtue of its heavy use of background structure. \cite{IshamPrima} (p. 17)
\end{quote}
What then is the difficulty?  According to Isham (and J. Butterfield), relativists worry about having two distinct causal structures, and  also wish to include nontrivial topologies to study various gravitational phenomena, including black holes, phase changes, and quantum cosmology \cite{IshamPrima,ButterfieldIsham}.
But there is no need to identify the two causal structures, even asymptotically, as long as the effective curved metric's null cone is consistent with the flat metric's null cone.  Above we have presented a formalism which plausibly ensures that the effective causal structure is consistent with the background one by construction, so Isham's first objection is answered.  The second objection simply presupposes a fundamentally geometrical view of gravitation and thus begs the question against the SRA.  It also appears to assume that black hole phenomena must be inconsistent with Minkowksi spacetime, but, as we have seen, even the interior and the region up to the true singularity fit within Minkowski spacetime.  Butterfield and Isham also refer to ``conjectures to the effect that a non-perturbative quantization of this system would lead to quantum fluctuations of the causal structure around a quantum-averaged background metric that is not the original Minkowskian metric.'' \cite{ButterfieldIsham} (p. 70)  Such conjectures, depending on the details, might be most welcome in the SRA, because fluctuations \emph{around} the flat background (as opposed to fluctuations \emph{bounded by} the flat background) would violate $\eta$-causality.

	It appears that the SRA gives a formalism in which the cost of including a flat background is low, while the benefits are noticeable.  As T. Thiemann notes, giving up Minkowski spacetime violates the Wightman axioms for quantum field theory \cite{Thiemann}.  Not everything is lost if one works on an \emph{arbitrary} background spacetime, but ``{\it the whole framework of ordinary quantum field
theory breaks down once we make the gravitational field (and the differentiable
manifold) dynamical, once there is no background metric any longer} !'' \cite{Thiemann} (emphasis in the original).  One could follow Thiemann's bold project of trying to do quantum field theory without a background metric.  Moreover, work is being done on implementing causality in such contexts \cite{MarkopoulouSmolin,Markopoulou,Livine}.  Even so, the task is sufficiently difficult that more conservative approaches to canonical quantization are worth exploring.  

We conclude that \emph{background-dependent nonperturbative} quantization, canonical or otherwise, of the SRA would be worthwhile. If one uses path integration, the $\eta$-stably causal configuration space might provide an appealing set of curved metrics over which to integrate to impose causality.  
 
	With the requirement of $\eta$-causality imposed, it follows that any ``SRA spacetime'' $(R^{4}, \eta_{\mu\nu}, g_{\mu\nu})$ is globally hyperbolic.  How so?  It follows from $\eta$-causality that the domain of dependence of an $\eta$-spacelike slice is in fact the whole of $R^{4}.$  But global 
hyperbolicity just is the possession of a Cauchy surface \cite{Wald}, so any $\eta$-causal SRA spacetime  $(R^{4}, \eta_{\mu\nu}, g_{\mu\nu})$  is globally hyperbolic.  Global hyperbolicity  avoids the Hawking black hole information loss paradox \cite{EarmanInfo}. 


\section{Conclusion}

	We have aimed to take special relativity seriously, including its causal, metrical, and topological structures, while viewing gravity as described by Einstein's field equations.  We found a good kinematical description of the relation between the null cones.  Plausibly one can deform any physically relevant solution into one in which the proper null cone relation obtains.  Having done so, one can adopt a new set of variables which ensure that the proper relation holds automatically.  Gauge transformations form not a group, but a groupoid.  As a result of using the flat metric, the problem of defining causality in quantum gravity is solved, so background-dependent nonperturbative quantization of the SRA might be worthwhile.  Furthermore, every SRA solution is globally hyperbolic. 


\section{Acknowledgments} 

This work is based on a portion of the first author's dissertation written at the University of Texas at Austin, supervised by the second author.  J. B. P. thanks M. Choptuik, J. L. Friedman, I. Goldman, S. Krasnikov, M. Visser, A. N. Petrov, Yu. M. Chugreev, J. M. M. Senovilla, A. A. Logunov, and E. Woolgar for discussions.


\bibliography{Pitts}

\begin{thebibliography}{10}

\bibitem{GuptaReview}
S.~N. Gupta, Rev. Mod. Phys. {\bf 29},  334  (1957).

\bibitem{Kraichnan}
R. Kraichnan, Physical Review {\bf 98},  1118  (1955).

\bibitem{Thirring}
W. Thirring, Annals of Physics (N.Y.) {\bf 16},  96  (1961).

\bibitem{NSSexlField}
O. Nachtmann, H. Schmidle, and R.~U. Sexl, Acta Phys. Aust. {\bf 29},  289
  (1968).

\bibitem{Huggins}
E.~R. Huggins, Ph.D. thesis, California Institute of Technology, 1962.

\bibitem{Feynman}
R.~P. Feynman, F. Morinigo, W. Wagner, and B. Hatfield, {\em Feynman Lectures
  on Gravitation} (California Institute of Technology, reprint Addison-Wesley
  (1995), Reading, Mass., 1963).

\bibitem{Weinberg64c}
S. Weinberg, Physica Letters {\bf 9},  357  (1964).

\bibitem{Weinberg64d}
S. Weinberg, Physical Review {\bf 135},  B1049  (1964).

\bibitem{WyssGC}
W. Wyss, Foundations of Physics Letters {\bf 10},  85  (1997).

\bibitem{OP}
V.~I. Ogievetsky and I.~V. Polubarinov, Annals of Physics (N.Y.) {\bf 35},  167
   (1965).

\bibitem{vanNieuwenhuizen}
P. van Nieuwenhuizen, Nuclear Physics B {\bf 60},  478  (1973).

\bibitem{Deser}
S. Deser, General Relativity and Gravitation {\bf 1},  9  (1970).

\bibitem{DeserQG}
D.~G. Boulware and S. Deser, Annals of Physics (N.Y.) {\bf 89},  193  (1975).

\bibitem{Fronsdal}
J. Fang and C. Fronsdal, Journal of Mathematical Physics {\bf 20},  2264
  (1979).

\bibitem{Cavalleri}
G. Cavalleri and G. Spinelli, Riv. Nuovo Cim. {\bf 3},  1  (1980).

\bibitem{DaviesFang}
P.~C.~W. Davies and J. Fang, Proceedings of the Royal Society (London) A {\bf
  381},  469  (1982).

\bibitem{Meszaros}
A. M\'{e}sz\'{a}ros, Acta Phys. Hung. {\bf 59},  379  (1986).

\bibitem{Grishchuk}
L.~P. Grishchuk, A.~N. Petrov, and A.~D. Popova, Communications in Mathematical
  Physics {\bf 94},  379  (1984).

\bibitem{Grishchuk90}
L.~P. Grishchuk, Soviet Physics Uspekhi {\bf 33},  669  (1990).

\bibitem{LogunovFund}
A.~A. Logunov and M.~A. Mestvirishvili, Theor. Math. Phys. {\bf 86},  1
  (1991).

\bibitem{LogunovBook}
A.~A. Logunov, {\em The Relativistic Theory of Gravitation} (Nova Science,
  Commack, N.Y., 1998).

\bibitem{SliBimGRG}
J.~B. Pitts and W.~C. Schieve, General Relativity and Gravitation {\bf 33},
  1319  (2001), gr-qc/0101058.

\bibitem{BoulangerEsole}
N. Boulanger and M. Esole, Classical and Quantum Gravity {\bf 19},  2107
  (2002), gr-qc/0110072.

\bibitem{NullCones}
J.~B. Pitts and W.~C. Schieve, www.arxiv.org {\bf gr-qc/0110004},    (2002).

\bibitem{PetrovHarmonic}
A.~N. Petrov, Astronomical and Astrophysical Transactions {\bf 1},  195
  (1992).

\bibitem{EarmanInfo}
G. Belot, J. Earman, and L. Ruetsche, Brit. J. Phil. Sci. {\bf 50},  189
  (1999).

\bibitem{IARD2002}
J.~B. Pitts and W.~C. Schieve, Foundations of Physics {\bf 33},  1315  (2003).

\bibitem{Penrose}
R. Penrose,  in {\em Essays in General Relativity---A Festschrift for Abraham
  Taub}, edited by F.~J. Tipler (Academic, New York, 1980).

\bibitem{SuperCens}
M. Visser, B.~A. Bassett, and S. Liberati, Nuclear Physics B (Proc. Suppl.)
  {\bf 88},  267  (2000), gr-qc/9810026.

\bibitem{PertSuperCens}
M. Visser, B. Bassett, and S. Liberati,  in {\em General Relativity and
  Relativistic Astrophysics: Eighth Canadian Conference, Montreal, Quebec,
  {J}une 1999}, edited by C.~P. Burgess and R.~C. Myers (AIP, Melville, N.Y.,
  1999), gr-qc/9908023.

\bibitem{PenroseWoolgar}
R. Penrose, R.~D. Sorkin, and E. Woolgar, www.arxiv.org {\bf gr-qc/9301015},
  (1993).

\bibitem{GaoWald}
S. Gao and R.~M. Wald, Classical and Quantum Gravity {\bf 17},  4999  (2000),
  gr-qc/0007021.

\bibitem{PalmerMarolf}
B.~C. Palmer and D. Marolf, Physical Review D {\bf 67},  044012  (2003),
  gr-qc/0211045.

\bibitem{Ohanian}
H. Ohanian and R. Ruffini, {\em Gravitation and Spacetime}, 2nd  ed. (Norton,
  New York, 1994).

\bibitem{LogunovBasic}
A.~A. Logunov, Theor. Math. Phys. {\bf 92},  826  (1992).

\bibitem{ChugreevCause}
Y.~V. Chugreev, Theoretical and Mathematical Physics {\bf 138},  293  (2004).

\bibitem{LoskutovRad96}
Y.~M. Loskutov, Theor. Math. Phys. {\bf 107},  686  (1996).

\bibitem{MTW}
C. Misner, K. Thorne, and J. Wheeler, {\em Gravitation} (Freeman, New York,
  1973).

\bibitem{Wald}
R. Wald, {\em General Relativity} (Univ. Chicago, Chicago, 1984).

\bibitem{AshtekarGeroch}
A. Ashtekar and R. Geroch, Reports on Progress in Physics {\bf 37},  1211
  (1974).

\bibitem{GerochSpinor}
R. Geroch, Journal of Mathematical Physics {\bf 9},  1739  (1968).

\bibitem{GotayDemaretPRD}
M.~J. Gotay and J. Demaret, Physical Review D {\bf 28},  2402  (1983).

\bibitem{GotayDemaretNPB}
M.~J. Gotay and J. Demaret, Nuclear Physics B (Proc. Suppl.) {\bf 57},  230
  (1997).

\bibitem{RosenSchwarzschild}
N. Rosen,  in {\em Relativity: Proceedings of the Relativity Conference in the
  Midwest, held at Cincinnati, Ohio, {J}une 2-6, 1969}, edited by M. Carmeli,
  S.~I. Fickler, and L. Witten (Plenum, New York, 1970), p.\ 229.

\bibitem{Bel}
L. Bel, Journal of Mathematical Physics {\bf 10},  1501  (1969).

\bibitem{JNW}
A.~I. Janis, E.~T. Newman, and J. Winicour, Physical Review Letters {\bf 20},
  878  (1968).

\bibitem{CooperstockJunevicus}
F.~I. Cooperstock and G.~J.~G. Junevicus, Nuovo Cimento B {\bf 16},  387
  (1973).

\bibitem{VlasovIncorrect}
A.~A. Vlasov, Soviet Journal of Nuclear Physics {\bf 51},  1148  (1990).

\bibitem{LoskutovSchwarzschild}
Y.~M. Loskutov, Soviet Journal of Nuclear Physics {\bf 54},  903  (1991).

\bibitem{Bicak}
J. Bi\v{c}\'{a}k,  in {\em Einstein Field Equations and Their Physical
  Implications--Selected essays in honour of Juergen Ehlers}, edited by B.~G.
  Schmidt (Springer, Berlin, 2000), gr-qc/0004016.

\bibitem{Goldstein}
H. Goldstein, {\em Classical Mechanics}, 2nd  ed. (Addison-Wesley, Reading,
  Mass., 1980).

\bibitem{Ionescu}
D. Ionescu, Theor. Math. Phys. {\bf 130},  287  (2002), gr-qc/0202058.

\bibitem{Griffiths}
D.~G. Griffiths, {\em Introduction to Electrodynamics}, 2nd  ed. (Prentice
  Hall, Englewood Cliffs, New Jersey, 1989).

\bibitem{HawkingChronology}
S.~W. Hawking, Physical Review D {\bf 46},  603  (1992).

\bibitem{Lehner}
L. Lehner, Classical and Quantum Gravity {\bf 18},  R25  (2001), gr-qc/0106072.

\bibitem{OdenReddy}
J.~T. Oden and J.~N. Reddy, {\em Variational Methods in Theoretical Mechanics}
  (Springer, Berlin, 1976).

\bibitem{Lacomba}
E.~A. Lacomba and W.~M. Tulczyjew, Journal of Mathematical Physics {\bf 23},
  2801  (1990).

\bibitem{GoldbergKlotz}
J.~N. Goldberg and F.~S. Klotz, International Journal of Theoretical Physics
  {\bf 7},  31  (1973).

\bibitem{IshamKakas}
C.~J. Isham and A.~C. Kakas, Classical and Quantum Gravity {\bf 1},  621, 633
  (1984).

\bibitem{Kuchar92}
K.~V. Kucha\v{r},  in {\em General Relativity and Gravitation 1992: Proceedings
  of the Thirteenth International Conference on General Relativity and
  Gravitation held at Cordoba, Argentina, 28 {J}une - 4 {J}uly 1992}, edited by
  R.~J. Gleiser, C.~N. Kozameh, and O.~M. Moreschi (Institute of Physics,
  Bristol, 1993).

\bibitem{Klauder2}
J.~R. Klauder, Journal of Mathematical Physics {\bf 42},  4440  (2001),
  gr-qc/0102041.

\bibitem{Eisenhart}
L.~P. Eisenhart, {\em Riemannian Geometry} (Princeton Univ., Princeton, 1926).

\bibitem{HallArab}
G.~S. Hall, Arab. J. Sci. Eng. {\bf 9},  87  (1984).

\bibitem{HallDiff}
G.~S. Hall,  in {\em Differential Geometry}, edited by W. Waliszewski, G.
  Andrzejczak, and P.~G. Walczak (PWN--Polish Scientific Publishers, Warsaw,
  1984).

\bibitem{HallNegm}
G.~S. Hall and D.~A. Negm, International Journal of Theoretical Physics {\bf
  25},  405  (1986).

\bibitem{Hall5d}
G.~S. Hall, M.~J. Rebou\c{c}as, J. Santos, and A.~F.~F. Teixeira, General
  Relativity and Gravitation {\bf 28},  1107  (1996).

\bibitem{BonaCollMorales}
C. Bona, B. Coll, and J.~A. Morales, Journal of Mathematical Physics {\bf 33},
  670  (1992).

\bibitem{Goldman}
I. Goldman, General Relativity and Gravitation {\bf 9},  575  (1978).

\bibitem{NCTRA}
J.~B. Pitts and W.~C. Schieve,  in {\em Relativistic Astrophysics: 20th Texas
  Symposium (December 10-15, 2000, Austin, Texas)}, edited by J.~C. Wheeler and
  H. Martel (AIP, Melville, New York, 2001), p.\ 763, gr-qc/0101097.

\bibitem{ChurchillVector}
R.~V. Churchill, Trans. Amer. Math. Soc. {\bf 34},  784  (1932).

\bibitem{Milnor}
J. Milnor,  in {\em Relativity, Groups, and Topology II}, edited by B.~S.
  DeWitt and R. Stora (Elsevier Science, New York, 1984).

\bibitem{Freifeld}
C. Freifeld,  in {\em Battelle Rencontres: 1967 Lectures in Mathematics and
  Physics}, edited by C. DeWitt and J.~A. Wheeler (Benjamin, New York, 1968).

\bibitem{ChurchillComplex}
R.~V. Churchill and J.~W. Brown, {\em Complex Variables and Applications},
  fifth ed. (McGraw-Hill, New York, 1990).

\bibitem{Israel}
W. Israel, {\em Differential Forms in General Relativity}, 2nd  ed. (Dublin
  Institute for Advanced Studies, Dublin, 1979).

\bibitem{GrishchukPetrov}
L.~P. Grishchuk and A.~N. Petrov, Soviet Physics JETP {\bf 65},  5  (1987).

\bibitem{PopovaPetrov}
A.~D. Popova and A.~N. Petrov, International Journal of Modern Physics A {\bf
  11},  2651  (1988).

\bibitem{Ramsay}
A. Ramsay, Advances in Mathematics {\bf 6},  253  (1971).

\bibitem{Hahn}
P. Hahn, Trans. Amer. Math. Soc. {\bf 242},  1  (1978).

\bibitem{Renault}
J. Renault, {\em A Groupoid Approach to $C^{*}$-algebras} (Springer, Berlin,
  1980).

\bibitem{Petrov}
A.~N. Petrov, International Journal of Modern Physics D {\bf 4},  451  (1995).

\bibitem{SenovillaCausal}
A. García-Parrado and J.~M.~M. Senovilla, Classical and Quantum Gravity {\bf
  20},  625  (2003), gr-qc/0207110.

\bibitem{IshamPrima}
C.~J. Isham,  in {\em Canonical Gravity: From Classical to Quantum.
  Proceedings, of the 117th WE Heraeus Seminar Held at Bad Honnef, Germany,
  13-17 September 1993}, edited by J. Ehlers and H. Friedrich (Springer,
  Berlin, 1994), gr-qc/9310031.

\bibitem{Solovev}
V.~O. Solov'ev, Theoretical and Mathematical Physics {\bf 65},  1240  (1986).

\bibitem{ButterfieldIsham}
J. Butterfield and C.~J. Isham,  in {\em Physics Meets Philosophy at the Planck
  Scale}, edited by C. Callender and N. Huggett (Cambridge Univ., Cambridge,
  2001), gr-qc/9903072.

\bibitem{Thiemann}
T. Thiemann, www.arxiv.org {\bf gr-qc/0110034},    (2001).

\bibitem{MarkopoulouSmolin}
F. Markopoulou and L. Smolin, Physical Review D {\bf 58},  084032  (1998),
  gr-qc/9712067.

\bibitem{Markopoulou}
F. Markopoulou, Classical and Quantum Gravity {\bf 17},  2059  (2000),
  hep-th/9904009.

\bibitem{Livine}
E.~R. Livine and D. Oriti, Nuclear Physics B {\bf 663},  231  (2003),
  gr-qc/0210064.

\end{thebibliography}
\bibliographystyle{prsty} 
\end{document}